\documentstyle[aps,graphicx,epsf]{revtex}
\tighten

\def \lleq {\lower0.9ex\hbox{ $\buildrel < \over \sim$} ~}
\def \ggeq {\lower0.9ex\hbox{ $\buildrel > \over \sim$} ~}
\newcommand{\sq}{\lower.25ex\hbox{\large$\Box$}}

\def\beq{\begin{equation}}
\def\eeq{\end{equation}}
\def\ber{\begin{eqnarray}}
\def\eer{\end{eqnarray}}

\begin{document}

\draft
\title{Inflation With Oscillations}
\topmargin -2cm
\author{M. Sami\\
Department of Physics, \\
Jamia Millia Islamia, New Delhi-110025, INDIA}
\maketitle
\begin{abstract}    
In this paper we investigate the general features of "Oscillatory
Inflation". In adiabatic approximation , we derive a general formula
for the number of e-foldings $\tilde{N}$ which reduces to the standard
expression in case of the slow role approximation and leads to the
Damour-Mukhanov type expression for the slowly varying adiabatic
index.  We apply our result to the logarithmic potential and arrive at
a simple and elegant formula for the number of e-foldings.We evolve
the field equations numerically and observe a remarkable agreement
with the analytical result.
\end{abstract}

\maketitle

\section{Introduction}
The inflationary universe scenario has become an integral part of the
standard model of universe. It does resolve the outstanding problems
of the standard model like the flatness problem, the horizon problem,
the homogeneity and isotropy problem etc. It also provides an
important clue for the origin of structure formation in the universe
[1,2, 3] .  The underlying idea of inflation is that there was an
epoch when the universe was slowly rolling down the flat wings of the
scalar potential such that vacuum energy was dominant leading to the
exponential growth of the scale factor. At the end of slow roll, the
universe falls into the core of the potential and fast oscillates near
the minimum leading to the re-heating ultimately. Damour and Mukhanov
have  discussed the possibility of inflation during the period
of oscillations [ 4 ] . Further discussion on the same theme can be
found in reference [ 5 ].  In fact, such a possibility was
first pointed out by Turner[ 6 ]. Inflation with oscillation can be
realized by a non-convex potential $V(\phi)$ ( $V_{,\phi\phi} < 0 $)
with small convex core near the minimum of the potential and with
large flat wings such that universe spends most of the time away from
the core and inflates.  In this paper we study general aspects of
inflation during oscillations .  We derive general expression for the
number of e-folds which reduce to the standard formula in the slow
role regime and leads to the Damour-Mukhanov type expression for slowly
varying adiabatic index. 
  We  apply our result to the logarithmic
potential and show that the model exhibits simple analytical solution which has
remarkable agreement with the numerical simulation.

\section{Evolution equations and criteria of inflation with
oscillations:} The field evolution equations in the Freedman cosmology
have the form
\begin{equation}
\frac{d}{dt} {\left( \frac{1}{2} \dot{\phi}^2 + V \right)} = -3H\dot{\phi}^2  
\end{equation}
\begin{equation}
H^2 = \frac{8\pi}{3M^2}\rho
\end{equation}
\begin{equation}
\dot{\rho} = -3H(\rho+p) = -3H\dot\phi^2
\end{equation}
\begin{equation}
\frac{\ddot{a}}{a}=-\frac{4\pi}{3M^2}(\rho+3p)
\end{equation}
where \begin{equation} \rho=\frac{1}{2}\dot{\phi}^2+V(\phi) 
\end{equation}
\begin{equation}
p=\frac{1}{2}\dot{\phi}^2-V(\phi)
\end{equation}
 Equations (3) and (4) are not independent of (1) and (2) but will be useful
 in view of forthcoming discussion. Equation (3) reveals that the energy density
 in scalar field $\phi$ decreases because of the red shifting away of the kinetic part
 $\frac{\dot{\phi}^2}{2}$ .
 In what follows  , we shall assume that the potential is even and has
minimum at $\phi=0$ . When the field  initially being displaced from the
minimum of the potential, rolls below its slow roll value, the coherence oscillation regime $\omega >> H$ commences. The evolution 
equation can then be approximately solved by separating the two times scales namely
the fast oscillation time scale and the longer expansion time scale. On the first
time scale, the Hubble expansion can be neglected, and one obtain $\phi$ as a
function of time;
\begin{equation}
t-t_0 = \pm \int{\frac{1}{\sqrt{2(V_m-V(\phi))}}} d\phi, 
\end{equation}
where $ \rho \equiv V_m \equiv V(\phi_m) $ ; $V_m $ being the maximum current value of the 
potential energy and $\phi_m$ being the field amplitude .
On the longer time scale $\rho$ and $\phi_m$ slowly decrease because 
Hubble damping term in equation ( 1 ) .The average adiabatic index $\gamma$ 
is defined as ,
\begin{equation}
\gamma =\left \langle \frac{\rho+p}{\rho}\right\rangle=\left \langle\frac{\dot{\phi}^2}{\rho}\right\rangle
\end{equation}
where$ < . >$
denotes the time average over one oscillation .
 Equation (4) then tells that expansion during oscillations would 
 continue ($\ddot{a}>0)$  if $\gamma < \frac{2}{3}.$ The adiabatic evolution 
 of $a(t)$ and $\rho$ is given by ,
 \begin{equation}
 a(t) \propto t^{\frac{2}{3\gamma}}
 \end{equation}
 \begin{equation} \rho \equiv V(\phi_m)\propto
 { t}^{-2} \end{equation}
As $\ddot{\phi}={-dV \over d\phi}$,the condition $\gamma<\frac{2}{3} $ can equivalently be written  [ 6 ] ,
$$\gamma =\left\langle \frac{\dot{\phi}^2}{\rho}\right\rangle = \frac{<\phi V_{,\phi}>}{V_m} = 2(1 -\frac{<V>}{V_m}) $$ 
\begin{equation}
= 2 \frac{\int^{\phi_m}_0(1-V(\phi)/V_m)^{\frac{1}{2}}d\phi}{ \int^{\phi_m}_0(1-V(\phi)/V_m)^{-\frac{1}{2}}d\phi}
< \frac{2}{3}
\end{equation} 
It is suggestive to write (11) in an equivalent form,
\begin{equation} 
\langle U(\phi)\rangle \equiv\left \langle V-\phi V_{,\phi}\right\rangle > 0
\end{equation} The expression (12) has a simple geometric meaning;
the average over one oscillation of the intercept U($\phi$) at $\phi$ should be positive
to ensure accelerated expansion.
 In models of inflation with oscillation the slow roll value of the
field $\phi_{s} $   is " far away" from convex core given by $\phi_{c}$ such that$ \langle U \rangle$ is
positive before $\phi$ rolls below certain   critical value of the field
$\phi_{cr}$ Fig.1 . As a result the universe undergoes accelerated
expansion  ($\ddot a > 0) .$ This is because the average of intercept
$ U(\phi)$ is negative for   $ 0 < \phi < \phi_{c} $  and the
positive contribution to $ <U(\phi)>$ starts coming only as $\phi$   gets bigger than
$ \phi_{c} $ . Obviously one has to take $ \phi $ further to some  $
\phi_{cr}$ in order to cancel the negative contribution to $ <U(\phi)>$
coming from the    convex core .
In certain sense the $\phi_{cr}$ defines the effective size of
convex core .   Clearly , as $\phi$ rolls below $\phi_{cr}$ , the
average of $ U(\phi)$ over one oscillation  turns negative and
consequently inflation ceases loosing to oscillations without
accelerated  expansion . Hence for inflation to occur during oscillations ,
the field amplitude $\phi_m$ should very between $\phi_{cr}$ and
$\phi_s$ . We shall call this regime " oscillatory   inflation regime " .
\begin{figure}[h]
\centering
\resizebox{!}{3.0in}{\includegraphics{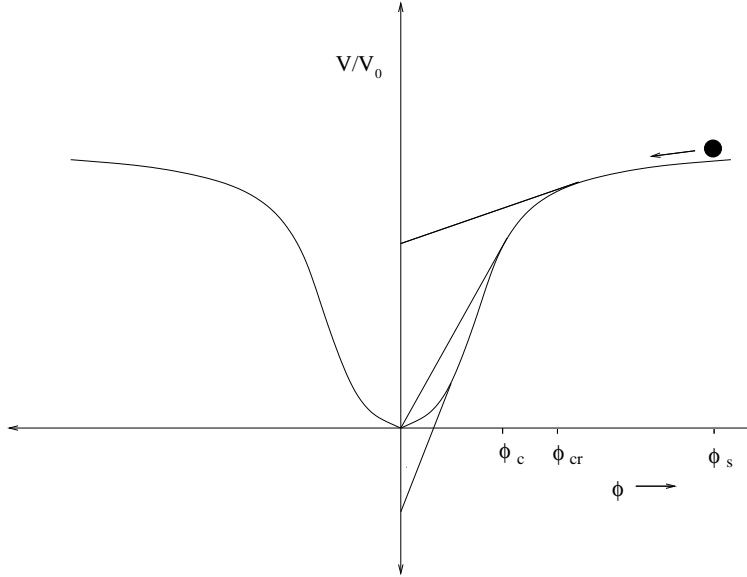}}
\caption{Qualitative picture of an effective potential
 showing the intercept $U(\phi)$ of the tangent to the curve
  $V(\phi)$ at various values of the field  $\phi .$For $\phi < \phi_c ,$the value
  of the intercept is negative . $U(\phi) $is positive for $\phi > \phi_c$ . 
  The critical field $\phi_{cr}$ is fixed such that the positive contribution coming
  from the  interval $ \phi_c < \phi < \phi_{cr}$ to the average 
  intercept $\langle U ( \phi ) \rangle $ exactly cancel the negative contribution
  of the convex core . $\langle U(\phi) \rangle$ is positive as $\phi$ varies between
 $ \phi_{cr}$ and $\phi_s$ and the universe expands in this regime called the
  "oscillatory inflation regime ". }
\end{figure}
 \section{Model:}
 We shall study here the logarithmic potential which may implement
inflation with oscillations,
\begin{equation}
{V}(\phi) =V_0 \ln\left[1+(\phi/\phi_c)^2)\right] 
\end{equation}
This is the q tending to zero limit of the potential suggested by Damour and Mukhanov, 
\begin{equation}
V(\phi)={V_0 \over q}\left[\left({\phi^2 \over \phi_c^2}+1\right)^{q/2}-1 \right]
\end{equation}
The  potential has a convex core near the bottom of
potential given by $\phi_c$ whereas $V_{,\phi\phi}$ is negative  away from the core. For $\phi>>\phi_c$ ,$V(\phi)=2V_0\ln(\phi/\phi_c)$ and consequently $\gamma\simeq 1/\ln(\phi_m/\phi_c)
\equiv 1/\ln(\beta)$ 
\subsection {critical value of field} 
 In order to compute the size of inflation , 
 it  would be
 necessary to determine $\langle U \rangle$ :
 \begin{equation}
 \langle U\rangle ={\int_0}^1({V(x,\beta)-x{V_{,x}}(x,\beta)})dx 
 \end{equation}
 where $\beta\equiv\phi_m/\phi_c$ , $x\equiv\phi/\phi_m$ and
  $V(x,\beta)=V_0\ln(1+\beta^2x^2)$ in case of the  potential given by ( 13 ).
 Since only the sigh of $ \langle U \rangle $ is important we have omitted several positive
  constants in the expression( 15 ) .
 The average intercept 
 $ \langle U \rangle $ is plotted in Fig. 2 in case of the logarithmic potential under consideration .The value of $\beta$ where $ \langle U \rangle $ changes sign precisely determines the critical value of the field or equivalently $\beta_{cr}$.For the logarithmic potential $\beta_{cr}$ turns out to be nearly equal to 3.2.
  
\begin{figure}[h]
\centering
\resizebox{!}{3.0in}{\includegraphics{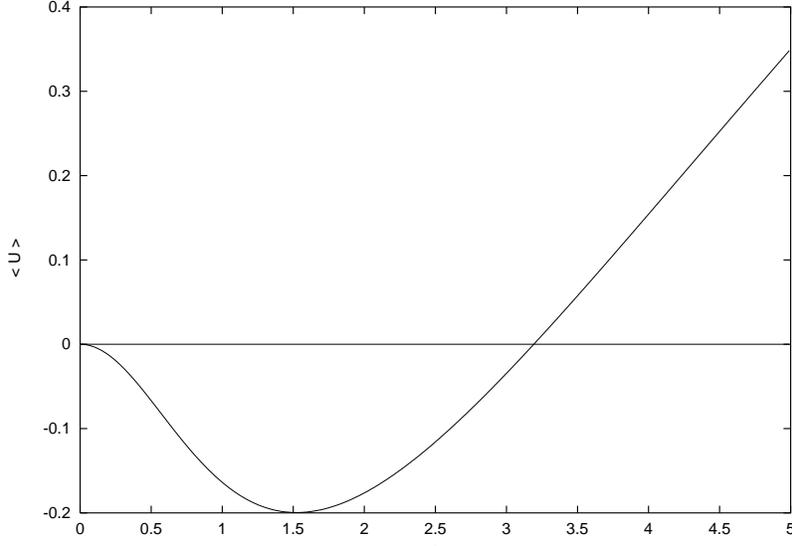}}
\caption{
Plot of the average intercept $\langle U(\phi) \rangle $ versus $\beta \equiv \phi_m/\phi_c $ 
for the potential given by Eq.( 13 ).$\langle U(\phi)\rangle $ changes sign as $\phi_m $ becomes
nearly equal to $3.2\phi_c$ indicating the beginning of oscillatory inflation regime.} 
\end{figure}
\subsection {slow roll field:}  The slow roll parameter $\epsilon$ is defined as :
\begin {equation}
  \epsilon = \frac{{M^2}}{16\pi} \left[\frac{V_{,\phi}}{V} \right]^2
  \end {equation}
  The slow roll ends when $\epsilon \simeq 1 .$ 
 Using ( 16 ) yields the slow roll value of the field $\phi{_s}$  ,
  \begin{equation}
  \sqrt{16 \pi}\times\beta_s\ln\beta_s\simeq M/\phi_c
  \end{equation}
   where $\beta_s = \frac{\phi_s}{\phi_c}$ and $\phi_s$ depends upon $\phi_c$ .It should be emphasized
   that $\beta_s$ or equivalently $\phi_s$ would always depend upon $\phi_c$ in case of
   non-power law potentials. 
   \subsection{Number of e-folds:} 
The number of e-folds is defined by ,
\begin{equation}
    N = \ln \frac{a_f}{a_i},
\end{equation}
 Since the scale factor "a" depends upon $\phi_m$ or equivalently upon $\beta,$ ( 18 ) should be used
 carefully . We divide the interval [$\beta_{cr},\beta_s$] into small slices ,
 $\Delta{\beta_j}=\beta_{j+1}-\beta_j $with $j=1,M$ such that $\beta_1=\beta_{cr}$ and
 $\beta_M=\beta_s.$The number of e-foldings obtained for the interval $\Delta{\beta_j}$ is given
 by ,
 $$N(\beta_j)\Delta{\beta}_j=-\ln\frac{a(\beta_{j+1})}{a(\beta_j)}$$ where $a_{j+1}<a_j$ for 
 universe rolling down from $\beta_s to \beta{cr}$
 using ( 9 ) and ( 10 ) leads to the following expression ,
 $$ N_j\Delta{\beta}_j={1\over3\gamma(\beta)}\ln\left({V(\beta_{j+1})/V(\beta_j)}\right)$$
 Expanding $V(\beta_{j+1})$  in Teller series in the neighborhood of $\beta_j$
 and using $\ln(1+x)\simeq{x}$ for $x<<1$ we get ,
 \begin{equation}
 N_j\Delta{\beta}_j=\left({1\over 3\gamma(\beta_j)} 
  \right){V'(\beta_j)\over V(\beta_j)}\Delta{\beta}_j
 \end{equation}
 Integrating ( 19 ) from $\beta_{cr}$ to $\beta_s$ one obtains the total number of 
 e-foldings ,
 $$ N=\int_{\beta_{cr}}^{\beta_s}{N(\beta)}d\beta$$ where
 $$ N(\beta) =\frac{V'(\beta)/V(\beta)}{3\gamma(\beta)}$$
$N(\beta)$ can be interpreted as the density of e-foldings with regard to $\beta$.
     The correct definition , however , should take into account the change in
     the comoving Hubble length [ 7 ] ,
     \begin{equation}
     \tilde{N}=\ln\left({a_fH_f \over a_iH_i}\right)
     \end{equation}
     Similar arguments as above can be used to obtain the expression
      for${\tilde{N}}$ ,
      \begin{equation}
     {\tilde{N}}=\int_{\beta_{cr}}^{\beta_s}{\tilde{N}(\beta)}d\beta
     \end{equation}
     where
     \begin{equation}
     \tilde{N}{(\beta)} ={\left(\frac{1}{3\gamma(\beta)}-{1\over 2}\right)}{V'(\beta)\over V(\beta)}
     \end{equation}
     Expression ( 22 ) formally written for oscillatory regime presents a general
     result for the number e-foldings . In fact ,in slow roll approximation ,
     $$\gamma={\dot{\phi}^2/2 -V\over\dot{\phi}^2/2+V}+1\simeq 
     {\dot{\phi}^2 \over V} \simeq\left({M^2\over24\pi}\right) \left
     ({V'\over V}\right)^2$$
     Substituting this expression in ( 22 ) leads to the standard formula for the
     number of e-foldings [ 2 ].Needless to say that $\tilde{N}$ nearly coincides
     with $N$ in case of the slow role regime .In case $\gamma$ is a slowly varying function of
     $\beta$, ${\tilde{N}}$ assumes the following form ,
     \begin{equation}
     {\tilde{N}} = {({1\over 3\gamma}-{1\over 2}}) \ln\left[\frac{V(\beta_{s})}{V(\beta_{cr})}\right]
    \end{equation}
    For a power law potential ( 23 ) readily reduces to the expression obtained by
    Liddle and Majumdar [ 6 ].
\section{Numerical simulation and analytical results}
We will show that a simple and elegant analytical formula for the number of e-folding can be 
obtained in case of the logarithmic potential ( 13 ). As  mentioned earlier
 that  $\gamma\simeq 1/\ln(\beta)$ in the present case [ 4,5 ].Substituting
the approximate value of $\gamma$ in ( 22 ) leads to a simple expression for $\tilde{N}(\beta)$,
\begin{equation}
\tilde{N}(\beta)={1\over3\beta}-{1\over2\beta \ln\beta}
\end{equation}
Using ( 21 ) we obtain the formula for the total number of e-foldings ,
\begin{equation}
{\tilde{N}}={1\over3} \ln\frac{\beta_s}{\beta_{cr}}-{1\over2}\ln\left(\frac{\ln\beta_s}{\ln \beta_{cr}}\right)
\end{equation}
It should be mentioned that we have used the approximate value of $\gamma$ to obtain the above expression for $\tilde{N}$.Since the approximation holds good away from the core and little contribution to the number of e-foldings comes from the regions near the core, we would expect that the expression ( 25 ) gives the correct value of  $\tilde{N}$.We have evolved the field equations numerically and studied the evolution of the number of e-foldings.  Fig2 displays the numerical evolution of
${\tilde{N}}$ for two different values of 
 $\phi_c$.The dependence of $\beta_s$ on  $\phi_c$  can in general
be red off from ( 17 ).The heights of maxima of $\tilde{N}$ are described by ( 25 ).We
see the remarkable agreement of our analytical result with the numerical
 simulation.As emphasized earlier [ 4, 5 ] the number of e-foldings $\tilde{N}$  
 is small.Even if $\phi_c$ is pushed to electro-weak scale 
$\tilde{N}$ does not exceed 10.One might think that brane world inflation
would be helpful here as the brane corrections enhances the prospects of
inflation [ 8, 9, 10]. Unfortunately , this is not so as the brane induced term increases the
Hubble expansion rate which is not relevant during oscillations. 
\begin{figure}[h]
\centering
\resizebox{!}{3.0in}{\includegraphics{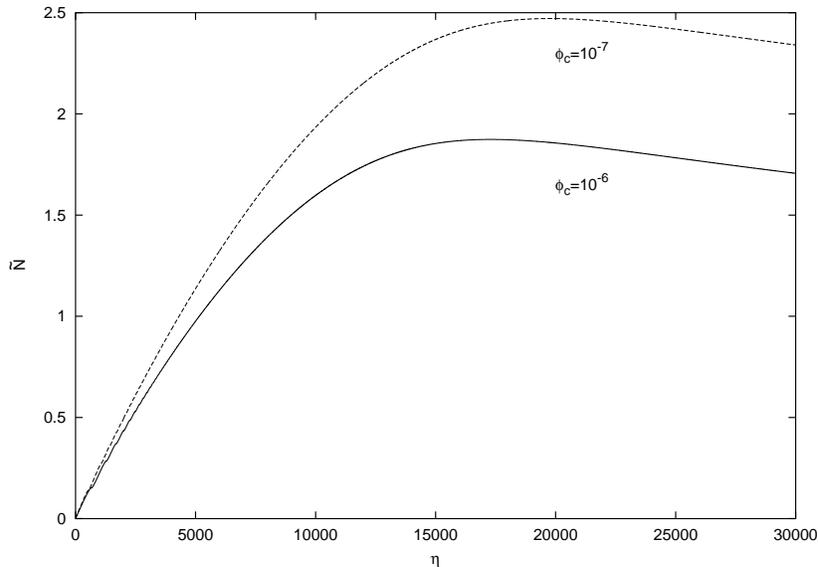}}
\caption{
Plot of $\tilde{N}$ versus $\eta \equiv tM $ based upon numerical simulation  
for two choices of $\phi_c$.The evolution takes place from the end of slow roll and counts
the number of e-foldings during the oscillations.The heights of maxima give the  actual number of
e-foldings described by ( 25 ).}
\end{figure}
We have described the general features of oscillatory inflation.We have been able
to obtain a  general expression for the number of e-foldings and demonstrated its
consistency. Applying the same to the logarithmic potential we obtained an elegant
and simple analytical formula for the number of e-foldings.We evolve the field equations numerically and observe a remarkable agreement
of the analytical result with the numerical simulation.
\section{Acknowledgments}
I am thankful to Varun Sahni and Tabish Qureshi for useful discussions.I also thankful to IUCAA where this work was started.

\end{document}